\documentclass[prl,twocolumn,showpacs]{revtex4}
\usepackage{graphicx}
\usepackage{textcomp}
\usepackage{amsmath}

\newcommand{\ket}[1]{\ensuremath{\left|\,#1\,\right\rangle}}                              
\newcommand{\eq}[2]{$#1$\,=\,$#2$}
\newcommand{\pe}{\mbox{$^\prime$}}

\begin{document}
\date{\today}
\author{S.~Kuhr}
\email{kuhr@iap.uni-bonn.de}
\author{W.~Alt}
\author{D.~Schrader}
\author{I.~Dotsenko}
\author{Y.~Miroshnychenko}
\author{W.~Rosenfeld}
\author{M.~Khudaverdyan}
\author{V.~Gomer}
\author{A.~Rauschenbeutel}
\author{D.~Meschede}
\affiliation{Institut f\"ur Angewandte Physik, Universit\"at
Bonn, Wegelerstr.~8, D-53115 Bonn, Germany}
\title{Coherence properties and quantum state transportation
in an optical conveyor belt}

\begin{abstract}
We have prepared and detected quantum coherences with long
dephasing times at the level of single trapped cesium atoms.
Controlled transport by an ``optical conveyor belt'' over
macroscopic distances preserves the atomic coherence with slight
reduction of coherence time. The limiting dephasing effects are
experimentally identified and are of technical rather than
fundamental nature. We present an analytical model of the
reversible and irreversible dephasing mechanisms. Coherent
quantum bit operations along with quantum state transport open
the route towards a ``quantum shift register'' of individual
neutral atoms.
\end{abstract}
\date{April 11, 2003}

\maketitle Obtaining full control of all internal and external
degrees of freedom of individual microscopic particles is the
goal of intense experimental efforts. The long lived internal
states of ions and neutral atoms are excellent candidates for
quantum bits (qubits), in which information is stored in a
coherent superposition of two quantum states. Once sufficient
control of individual particles is established, one of the most
attractive perspectives is the engineered construction of quantum
systems of two or more particles. Their coherent interaction is a
key \mbox{element} for the realization of quantum gates and can
be implemented by transporting selected qubits into an
interaction zone \cite{Kielpinski02}. In this context, ions have
successfully been transported between distinct locations while
maintaining internal-state coherence \cite{Rowe02}.

In this letter we report on the coherence properties and on a
quantum state transportation of neutral atoms in a standing wave
dipole trap. We present a scheme of preparation and detection of
the electronic hyperfine ground states, which has been applied to
ensembles as well as to single atoms. Along with the
demonstration of 1-qubit-rotations, our system is a promising
candidate for storing quantum information, since the hyperfine
ground states exhibit long ($\sim$200 ms) coherence times. In
particular, the coherence even persists while transporting the
atoms over macroscopic distances. This opens a route towards
quantum gates via cavity-mediated atom--atom coupling.


We trap cesium atoms in a standing wave dipole trap
(\eq{\lambda}{1064}\,nm) with a potential depth of
\eq{U_0}{1}\,mK, loaded from a high-gradient magneto-optical trap
(MOT) \cite{Kuhr01,Schrader01}. The single-atom transfer
efficiency between the two traps is better than 95\%
\cite{Frese00}. The MOT is also used to determine the exact
number of trapped atoms by observing their fluorescence. By
shifting the standing wave pattern along the direction of beam
propagation, we can transport the atoms over millimeter-scale
distances. This is realized by mutually detuning the frequencies
of the dipole trap laser beams with acousto-optical modulators.
Additionally, we use microwave radiation at \eq{\omega_{\rm
hfs}/2\pi}{9.2}\,GHz to coherently drive the
$\ket{\mbox{\eq{F}{4},
\eq{m_F}{0}}}\rightarrow\,\ket{\mbox{\eq{F}{3}, \eq{m_F}{0}}}$
clock transition of the $6^2S_{1/2}$ ground state with Rabi
frequencies of \eq{\Omega/2\pi}{10}\,kHz.

The initial state is prepared by optically pumping the atom into
$\ket{\mbox{\eq{F}{4}, \eq{m_F}{0}}}$. For this purpose, we use a
$\pi$-polarized laser resonant with the
\eq{F}{4}\,$\rightarrow$\,\eq{F'}{4} transition and a repumping
laser on the \eq{F}{3}\,$\rightarrow$\eq{F'}{4} transition in a
magnetic guiding field of \eq{B}{100}\,\textmu T. A state
selective detection method, which discriminates between the
atomic hyperfine states, is implemented by exposing the atom to a
$\sigma^+$-polarized laser, resonant with the
\eq{F}{4}\,$\rightarrow$\eq{F'}{5} cycling transition. It pushes
any atom in \eq{F}{4} out of the dipole trap, whereas an atom in
\eq{F}{3} remains trapped. The number of atoms remaining in the
dipole trap is then determined by transferring them back into the
MOT. In all experiments presented in this letter we thus compare
the number of atoms in the MOT before and after any coherent
manipulation in the dipole trap.

It is essential to push an atom in \eq{F}{4} out of the dipole
trap before it spontaneously decays into \eq{F}{3} due to
off-resonant excitation of the \eq{F'}{4} level. For this
purpose, the dipole trap is adiabatically lowered to 0.1\,mK
before the push-out laser is shined in perpendicular to the trap
axis with high intensity (\eq{I/I_0}{100}, where $I_0$ is the
saturation intensity). In this regime the resonant radiation
pressure force is stronger than the dipole force in radial
direction. We thus push the atom out of the trap within much less
than a quarter of the radial oscillation period ($\sim\,$1\,ms).
As a result, less than 1\% of the atoms prepared in \eq{F}{4}
survive the application of the push-out laser, whereas more than
95\% of the atoms in the \eq{F}{3} state remain trapped.

\begin{figure}
\begin{center}
  \includegraphics[width=\columnwidth]{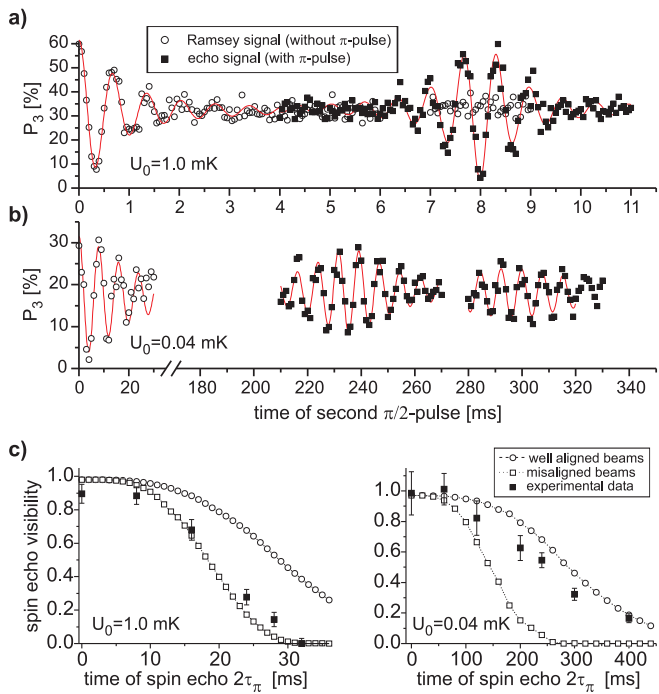}
\end{center}
\vspace{-0.3cm} \caption{Ramsey and spin echo signals. (a)
\eq{U_0}{1.0}\,mK. The application of a $\pi$-pulse at
\eq{t}{4}\,ms leads to a rephasing of the magnetic dipole moments
at \eq{t}{8}\,ms (squares). The hollow circles show a Ramsey
signal, obtained without a $\pi$-pulse, but with otherwise
identical parameters. Each point results from 30 shots with about
50 initial atoms. The lines are fits according to the analytic
model presented in the text. (b)~Ramsey signal and two different
spin echoes for \eq{U_0}{0.04}\,mK.  (c)~Decay of the spin echo
visibility for different trap depths (filled squares) together
with best (empty circles) and worst case predictions (empty
squares).} \label{fig:SpinEchoes}
\end{figure}

The coherence time between the $\ket{\mbox{\eq{F}{4},
\eq{m_F}{0}}}$ and the $\ket{\mbox{\eq{F}{3}, \eq{m_F}{0}}}$
state is measured using Ramsey's method of separated oscillatory
fields \cite{Ramsey}. We describe the coherent evolution of the
system by the semiclassical Bloch vector model
\cite{AllenEberly}. The initial state $\ket{\mbox{\eq{F}{4},
\eq{m_F}{0}}}$ corresponds to the Bloch vector
\eq{(u,v,w)}{(0,0,-1)}. The application of two $\pi/2$-pulses,
separated by a time interval $t$, leads to the well-known Ramsey
fringes $w(t)=\cos\delta t$, where $\delta=\omega-\omega_{\rm
hfs}$ is the detuning of the microwave frequency $\omega$ from
the atomic resonance $\omega_{\rm hfs}$. This leads to the
theoretical population transfer to \eq{F}{3},
\eq{P_3(t)}{[w(t)+1]/2}. In the experiment, $P_3$ is measured
using our state selective detection method. We count the number
of atoms in \eq{F}{3} and divide by the initial number of atoms
in the MOT. Since the atoms are trapped in separate potential
wells of the dipole trap and interactions between them can be
neglected, we perform our experiments with about 50~atoms in
order to reduce the measurement time. As there is no dependence
on the number of atoms the ensemble results can be transferred to
the single atom case.

The observed Ramsey signal decays due to inhomogeneous broadening
of the atomic resonance frequencies, see
Fig.~\ref{fig:SpinEchoes}. This frequency spread arises from the
energy distribution of the atoms in the dipole trap, which leads
to a corresponding distribution of light shifts. For an atom in
the bottom of a potential well (energy \eq{E}{0}), the clock
transition is light shifted by \eq{\hbar\delta_{\rm 0}}{\eta
U_0}, with \eq{\eta}{\omega_{\rm
hfs}/\Delta}\,=\,1.45\,$\times$\,10$^{-4}$ and $\delta_0\! <\! 0$
\cite{Kaplan02}. Here, \eq{\Delta/2\pi}{-64}\,THz is the
effective detuning of the dipole trap laser, taking into account
contributions of the D$_1$- and the D$_2$-line. In a potential of
\eq{U_0}{1}\,mK this differential light shift is \eq{\delta_{\rm
0}/2\pi}{-3.0}\,kHz. If the atom possesses an energy $E$\,$>$\,0
in the trapping potential, the average light shift decreases to
\eq{\delta_{\rm ls}(E)}{\delta_0+{\eta E}/(2\hbar)} in harmonic
approximation. In our experiment, the energy distribution of the
atoms has been measured to be a 3D Maxwell-Boltzmann distribution
with temperature $T$ \cite{Alt02b}. This results in a
corresponding distribution of differential light shifts,
\begin{equation}
    \widetilde{\alpha}(\delta_{\rm ls})=
\frac{2K^{3/2}}{\sqrt{\pi}}\sqrt{\delta_{\rm ls}-\delta_{\rm 0}}
\exp{[-K(\delta_{\rm ls}-\delta_{\rm 0})]},
\end{equation} with
$K$=${2\hbar}/({\eta k_{\rm B}T})$.  From this distribution, we
derive an analytic expression for the shape of the Ramsey signal.
For an atom with light shift $\delta_{\rm ls}$, the $w$-component
of the Bloch vector, after application of the two $\pi/2$-pulses,
is given by \eq{w_{\rm Ramsey}(t)}{\cos[(\delta+\delta_{\rm ls})
t]}. Averaged over a thermal ensemble, the Ramsey signal is the
integral over all light shifts, \eq{w_{\rm
Ramsey,inh}(t)}{\int_{\delta_0}^\infty w_{\rm
Ramsey}(t)\,\widetilde{\alpha}(\delta_{\rm ls})\,d\delta_{\rm
ls}}. The envelope of the Ramsey fringes is thus the Fourier
cosine transform $\alpha(t)$ of the distribution of light shifts,
yielding
\begin{eqnarray}
  w_{\rm Ramsey, inh}(t)&=&\alpha(t)\cos{[(\delta+\delta_0)
  t]},\\
\mbox{with}\
\alpha(t)&=& \left[1+2.79\left(t/T_2^*\right)^2\right]^{-3/4}.
\end{eqnarray} Despite this non-exponential decay, we have defined the
inhomogeneous or reversible dephasing time \eq{T_2^*}{1.67\,K} as
the 1/$e$-time of the amplitude $\alpha(t)$.

The Ramsey fringes presented in Figs.~\ref{fig:SpinEchoes}(a) and
(b) are fitted according to this model and yield
\eq{T_2^*}{(0.86\pm0.05)}\,ms and \eq{T_2^*}{(18.9\pm1.7)}\,ms
for \eq{U_0}{1.0}\,mK and \eq{U_0}{0.04}\,mK, respectively. The
maximum value of $P_3$ of 60\% in Fig.~\ref{fig:SpinEchoes}(a)
includes imperfections in the optical pumping process (-20\%) and
atom losses caused by inelastic collisions during the transfer
from the MOT  into the dipole trap (-20\%). The additional
reduction to 30\% in Fig.~\ref{fig:SpinEchoes}(b) occurs during
the lowering of the trap, where hot atoms are lost. Note that the
coherence is not impaired by these losses as the maximum
visibility is close to 100\% (see Fig~\ref{fig:SpinEchoes}).

The inhomogeneous dephasing can be reversed by applying a
$\pi$-pulse at \eq{t}{\tau_{\pi}} between the $\pi/2$-pulses, in
analogy to the spin-echo technique in nuclear magnetic resonance
\cite{Hahn50}. In optical dipole traps, this technique was
recently employed by another group \cite{Andersen03},
independently of our work. Our spin-echo signals are presented in
Figs.~\ref{fig:SpinEchoes}(a) and (b), showing that the
inhomogeneous dephasing can be reversed almost completely. In the
low dipole trap (Fig.~\ref{fig:SpinEchoes}(b)) we observe a
pronounced spin echo even for pulse delays of up to
\eq{2\tau_\pi}{300}\,ms. Similar to the Ramsey fringes, the shape
of the echo signal is
\begin{equation}
    w_{\rm echo}(t) = - \alpha(t-2\tau_\pi)
\cos{\left[(\delta+\delta_0)(t-2\tau_\pi)\right]}.
\end{equation}

In Fig.~\ref{fig:SpinEchoes}(c) we plot the visibility of the
echo signals as a function of $2\tau_\pi$ for the two trap
depths. The decay of the echo visibility is due to an
irreversible dephasing of the atomic coherence with a time
constant $T_2'$. We therefore extend our model to the case of a
time-varying detuning, $\delta(t)$, in order to account for a
variation of the precession angles of the Bloch vector,
\eq{\phi_1}{\int_0^{\tau_\pi}\delta(t)\,dt} and
\eq{\phi_2}{\int_{\tau_\pi}^{2\tau_\pi}\delta(t)\,dt}, before and
after the $\pi$-pulse. The phase difference $\phi_2-\phi_1$ is
expressed as a difference of the detuning, $\Delta\delta$,
averaged over $\tau_\pi$. The $w$-component of the Bloch vector
at \eq{t}{2\tau_\pi} then reads $w_{\rm
echo}(\Delta\delta,2\tau_\pi)=-\cos(\Delta\delta\,\tau_\pi)$. In
the following considerations, we assume that all atoms experience
almost the same fluctuation $\Delta\delta$, regardless of their
energy in the trap, i.\,e.~we consider a {\it homogeneous}
broadening effect. We obtain the signal from many repetitions of
the same experiment by integrating $w_{\rm echo}(\Delta\delta,
2\tau_\pi)$ over all fluctuation amplitudes $\Delta\delta$,
weighted by a probability distribution $p(\Delta\delta,\tau_\pi)$
which is assumed to be a Gaussian,
\begin{equation}
    p(\Delta\delta,\tau_\pi)=\frac{1}{\sqrt{2\pi}\,\sigma(\tau_\pi)}
    \exp\left[-\frac{(\Delta\delta)^2}{2\sigma(\tau_\pi)^2}\right]
\end{equation}
with mean $\overline{\Delta\delta}=0$ and variance
$\sigma(\tau_\pi)^2$. This results in an echo visibility of
\begin{equation}
 V(2\tau_\pi) = V_0\exp\left[-\tfrac{1}{2}\tau_\pi^2\sigma(\tau_\pi)^2\right].
\end{equation}
We found that the pointing instabilities of the dipole trap laser
beams are the dominant source of irreversible dephasing. Any
change of the relative position of the two interfering laser
beams changes the interference contrast and hence the light shift
$\delta_0$. Position fluctuations can arise from shifts of the
laser beam itself, from acoustic vibrations of the mirrors or
from air flow. Other dephasing effects such as intensity- and
magnetic field fluctuations, elastic collisions, heating,
fluctuations of the microwave power and the pulse duration were
also analyzed and found to be negligible compared to the effect
of the pointing instabilities.

The fluctuation of the trap depth due to pointing instabilities
was measured by mutually detuning the two dipole trap beams by
\eq{\Delta\nu}{10}\,MHz and overlapping them on a fast
photodiode. The amplitude of the resulting beat signal directly
measures the interference contrast of the two beams and is thus
proportional to the depth of the potential wells of the standing
wave dipole trap.  We found that the relative fluctuations for
time scales of $t$\,$>$\,100\,ms reach up to 3\%. From records of
the beat signal amplitude we extract the Allan variance, defined
as $\widetilde\sigma_{\rm A}^2(\tau) = \frac{1}{m}\sum_{k=1}^m
(\overline{x}_{\tau,k+1}-\overline{x}_{\tau,k})^2/2$, where
$\overline{x}_{\tau, k}$ denotes the average of the normalized
beat signal amplitude over the $k$-th time interval $\tau$
\cite{Allan66}. The Allan deviation $\widetilde\sigma_{\rm
A}(\tau_\pi)$ directly measures the difference of the detunings,
averaged over $\tau_\pi$, before and after the $\pi$-pulse. This
results in frequency fluctuations
\eq{\sigma(\tau_\pi)}{\sqrt{2}\,\delta_0\widetilde\sigma_{\rm
A}(\tau_\pi)}, yielding a theoretical spin echo visibility
\eq{V^{\rm th}(2\tau_\pi)}{\exp[-\widetilde\sigma_{\rm
A}(\tau_\pi)^2\delta_0^2\tau_\pi^2]} which is plotted together
with the measured spin echo  visibility in
Fig.~\ref{fig:SpinEchoes}(c). The upper curves show the expected
visibility in the case of well overlapped beams, whereas for the
lower curves, we slightly misaligned the beams so that variations
of the relative beam position cause a first order variation of
the beat signal amplitude, since the beams overlap on the slopes
of the Gaussian profile. Our datapoints lie in between these best
and worst case predictions.

The population decay time, $T_1$, can be neglected in our
experiment. It is governed by the scattering of photons from the
dipole trap laser, which couples the two hyperfine ground states
via a two-photon Raman transition. However, this effect is
suppressed due to a destructive interference effect
\cite{Frese00,Cline94} causing a relaxation on a timescale of
8\,s for \eq{U_0}{1}\,mK. As a consequence, the various
decay/dephasing mechanisms can be treated independently because
of their different time scales
($T_2^*$$\,\ll\,$$T_2\pe$$\,\ll\,$$T_1$).
\begin{figure}[!b]
\begin{center}
 \includegraphics[width=\columnwidth]{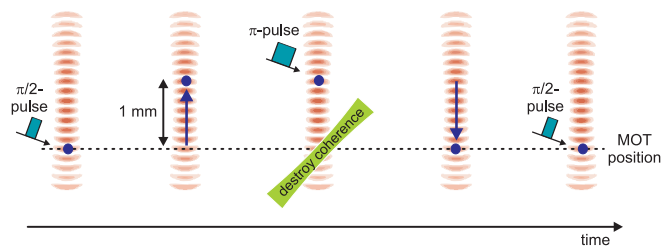}
\end{center}
\vspace{-0.5cm} \caption{Quantum state transportation. An atom
prepared in a superposition of hyperfine states is displaced by
1\,mm. Synchronously to the $\pi$-pulse, we shine in a state
mixing laser across the initial position. After transporting the
atom back to its initial position, the state superposition is
analyzed by means of a second $\pi/2$-pulse.}
\label{fig:SchemeCoherentTransport}
\end{figure}

 \begin{figure}
 \begin{center}
 \includegraphics[width=\columnwidth]{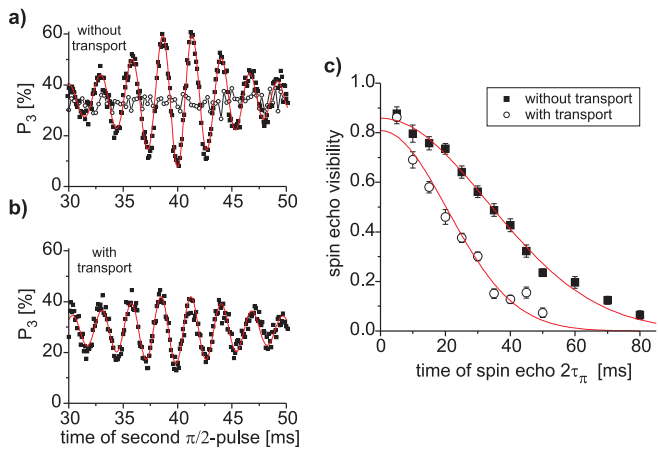}
\end{center}
\vspace{-0.5cm} \caption{Spin echo with and without transport,
according to the scheme in
Fig.~\ref{fig:SchemeCoherentTransport}. (a)~Without transport,
with (circles) and without mixing laser (squares); (b)~Atoms are
transported by 1\,mm, with mixing laser applied at the initial
position. The lines in (a) and (b) are fits according to the
analytic model presented in the text. (c)~Visibility of the spin
echo. The faster decay of the visibility with transport can be
explained by heating of the atoms during the transportation
procedure. }
  \label{fig:SpinEchoTransport}
\end{figure}

Finally, we demonstrate the controlled quantum state
transportation of neutral atoms. We show that the atomic
coherence persists while moving the atoms back and forth over
macroscopic distances by shifting the standing wave dipole trap.
For this, we essentially perform a spin echo measurement, with
the addition that the atoms are transported between the microwave
pulses. The sequence is visualized in
Fig.~\ref{fig:SchemeCoherentTransport}. After the $\pi/2$-pulse
at the MOT position, the atom is displaced by 1\,mm before the
$\pi$-pulse is applied. The atom is then transported back to its
initial position where we apply the second $\pi/2$-pulse. The
corresponding spin echo signal is shown in
Fig.~\ref{fig:SpinEchoTransport}, together with a reference
signal, showing a spin echo obtained without transportation. The
spin echo prevails if we transport the atom between the microwave
pulses, however with slightly reduced visibility, see
Fig.~\ref{fig:SpinEchoTransport}(b). In these experiments, the
dipole trap is lowered to \eq{U_0}{0.1}\,mK in order to guarantee
long coherence times along with high transportation efficiencies.
The atoms were transported over a distance of 1\,mm within 2\,ms,
resulting in an acceleration of
\eq{a}{1.0}\,$\times$\,10$^{3}$\,m/s$^2$.

In order to underline that the coherence has actually been moved
spatially in this experiment, we illuminate the position of the
MOT with an off-resonant ``state mixing laser'' (detuning
+30\,GHz) simultaneously with the $\pi$-pulse for 3~ms. The
parameters of the state mixing laser, which is focused to a waist
of 50\,\textmu m, are chosen such that it incoherently mixes the
two hyperfine states. A scattering rate of 2~photons/ms is
sufficiently high to just mix the hyperfine states but at the
same time to minimize the influence on the transported atoms.
Indeed the fringes disappear when the atoms are not transported.
The coherence time is reduced by a factor of two by the
transportation procedure. This can be explained by the abrupt
acceleration of the potential \cite{Schrader01}. For each of the
two transports, the acceleration is first abruptly changed from
zero to $a$, from $a$ to $-a$ after half the transportation
distance, and back to zero at the end. In order to calculate the
heating effect caused by this abrupt acceleration, we assume an
initial energy of \eq{E}{0.3\,U_0}, calculated from the
$T_2^*$-times of the signals of Fig.~\ref{fig:SpinEchoTransport}.
Numerical calculations reveal that the maximum energy gain caused
by the 6 non-adiabatic changes of the acceleration amounts to
\eq{\Delta E}{0.15\,U_0}\,(=\,150\,\textmu K). This energy gain
causes a change of the average detuning $\Delta\delta$, large
enough to account for the observed faster decay of the visibility
of the spin echo. The existence of a heating effect is supported
by the measurement of an atom survival probability with
transportation of 70\%, compared to 80\% without transportation.

Using sodium atoms in a red detuned Nd:YAG dipole trap
(\eq{U_0}{0.4}\,mK), a dephasing time of \eq{T_2^*}{15}\,ms was
reported \cite{Davidson95}, which is comparable to our
observation. In the same paper, significantly longer coherence
times (\eq{T_2^*}{4}\,s) were observed in blue detuned traps
\cite{Davidson95}. In other experiments, the inhomogeneous
broadening has been reduced by the addition of a weak light
field, spatially overlapped with the trapping laser field and
whose frequency is tuned in between the two hyperfine levels
\cite{Kaplan02}. Due to the different wavelengths the application
of this technique is impossible in a standing wave dipole trap.
Of course, cooling the atoms to the lowest vibrational level,
e.\,g.~by using Raman sideband cooling techniques, would also
reduce inhomogeneous broadening.

Our experiment opens the route to the realization of a ``quantum
shift register''. Most recently, we spatially resolved single
atoms in separate potential wells of the dipole trap by means of
an intensified CCD camera. This fact could allow us to read out
the quantum states of the trapped atoms individually. Moreover,
we plan to individually address the atoms in a magnetic gradient
field. Finally, the possibility of coherently transporting
quantum states should allow us to let atoms interact at a
location different from the preparation and read out. More
specifically, our experiments aim at the deterministic transport
of two or more atoms into an optical high finesse resonator,
where they could controllably interact via photon exchange. This
should enable us to entangle neutral atoms \cite{Zheng00} or to
realize quantum gates \cite{Pellizzari95}. This letter
demonstrates that in these experiments state preparation and
detection, as well as quantum state rotations can now take place
outside the cavity.

This work was supported by the Deutsche Forschungsgemeinschaft
and the EC. 

\end{document}